\documentclass[12pt,a4paper]{article}
\usepackage{axodraw}
\usepackage{cite}

\def\bea{\begin{eqnarray}}  
\def\eea{\end{eqnarray}}  
\def\bc{\begin{center}}
\def\ec{\end{center}}

\begin{document}
\pagestyle{empty}
\begin{flushright}
IFT-6/2007\\
{\tt hep-ph/yymmnnn}\\
{\bf \today}
\end{flushright}
\vspace*{5mm}
\begin{center}

{\large {\bf Violation of the Appelquist-Carazzone decoupling in non-SUSY GUT}}
\vspace*{1cm}

{\bf Piotr~H.~Chankowski} and {\bf Jakub Wagner}
\vspace{0.5cm}
 
Institute of Theoretical Physics, Warsaw University, Ho\.za 69, 00-681,
Warsaw, Poland

\vspace*{1.7cm}
{\bf Abstract}
\end{center}
\vspace*{5mm}
\noindent
{We point out that in non supersymmetric GUTs, in which the $SU(5)$ gauge 
symmetry is broken down to the Standard Model gauge group by a $\mathbf{24}$ 
Higgs multiplet the Appelquist-Carazzone decoupling is violated. This is 
because the $SU(2)_L$ Higgs triplet contained in the $\mathbf{24}$ acquires 
a dimensionfull coupling to the $SU(2)_L$ Higgs doublets which is proportional 
to the GUT breaking vacuum expectation value (VEV) $V$. As a result, at one 
loop heavy gauge and Higgs fields contribution to tadpoles generate a VEV of 
the triplet which is not suppressed for $V\rightarrow\infty$ and violates 
the custodial symmetry.
}
\vspace*{1.0cm}
\date{\today}


\vspace*{0.2cm}
 
\vfill\eject
\newpage

\setcounter{page}{1}
\pagestyle{plain}

\section{Introduction}

Grand Unified Theories (GUTs) typically predict the existence of many 
new superheavy particles. Even in the simplest $SU(5)$ model of Georgi 
and Glashow \cite{GEGL}, in addition to the superheavy gauge bosons, 
there are two multiplets of the Higgs fields with masses of the order 
of the GUT scale. Commonly the Appelquist-Carazzone principle \cite{APPCAR} 
is invoked to argue that in the low energy observables like $M_W$, $M_Z$, 
$\rho$, etc. all physical effects (as opposed to those which merely 
renormalize the couplings) of superheavy particles are suppressed by 
inverse powers of the GUT scale, i.e. that these particles decouple.

Nonsupersymmetric GUT models are known to suffer from the hierarchy problem
\cite{GILD} whose essence is that the separation of the GUT and the electroweak
scales is unnatural and requires an extremely precise fine-tuning of the GUT 
model parameters. However, the fine-tuning is a problem only if one wants to 
predict the electroweak scale in terms of the original parameters of the GUT 
model. It is not seen in practical calculations of electroweak observables 
if the electroweak scale itself is taken as one of the input observables. 
Thus, from the practical point of view, the fine-tuning can be ignored and one 
believes that predictions for low energy electroweak observables 
calculated in the GUT model should coincide, up to suppressed terms, with  
predictions obtained in the Standard Model (SM).\footnote{Our considerations
will be purely theoretical. Therefore we ignore the fact that in reality the 
three gauge coupling constants of the SM do not unify, that is, the strong 
coupling constant predicted by simple nonsupersymmetric GUTs disagrees with 
the measured one.}
Verification of this hypothesis by explicit calculation of 
a particular electroweak observable undertaken long time ago
by Senjanovic and Sokorac \cite{SENSOK} seemed to confirm this expectations. 
The heuristic explanation given by these authors (but attributed to F.
Wilczek) is that the decoupling is ensured by the gauge invariance: 
once the heavy fields are removed from the GUT Lagrangian, it possesses the
unbroken $SU(3)\times SU(2)_L\times U(1)_Y$ gauge symmetry. 

However our recent result indicates that the explanation offered by
Senjanovic Sokorac and Wilczek may not always be true. In \cite{CHPOWA2} 
we have considered the SM supplemented with an $Y=0$ (where $Y$ is the 
hypercharge) $SU(2)_L$ triplet $\phi$ and found that its effects on electroweak
observables are not suppressed for infinitely heavy additional (one 
neutral and one charged) scalars, if the dimensionfull coefficient $\mu$ of 
the coupling $H^\dagger_{\rm SM}\phi H_{\rm SM}$ (where $H_{\rm SM}$ is the 
SM $SU(2)_L$ doublet) is of order of the masses of these scalars. 
This violation of the Appelquist-Carazzone decoupling (found also in 
\cite{DAWSON,CDK}) can be attributed \cite{CHPOWA2} to the fact that once 
the electroweak symmetry is broken, the triplet is not protected against 
acquiring radiatively a VEV. In electroweak observables
this effect enters via a nonsuppressed contribution of tadpoles which can be 
absorbed into the redefinition of the triplet VEV. As a result, although
the tree level triplet VEV vanishes as the masses of the additional scalars
grow, the effective triplet VEV does not, causing a nonnegligible violation
of the familiar custodial symmetry protecting the $\rho$ parameter.

The triplet model considered in \cite{CHPOWA2} constitutes a clear 
counterexample to the explanation of Senjanovic, Sokorac and Wilczek: 
before the electroweak symmetry is broken one can remove the whole 
triplet $\phi$ from the Lagrangian without spoiling its full
$SU(2)_L\times U(1)_Y$ gauge invariance.
Yet, the explicit complete calculation of the $\rho$ parameter demonstrates
that the decoupling is violated. Crudely speaking, removing the heavy 
degrees of freedom from the Lagrangian does not necessarily commute
with the breaking of the electroweak symmetry. 

An immediate consequence of these result is that the Appelquist-Carazzone 
decoupling should also be violated in nonsupersymmetric GUTs\footnote{We 
comment on supersymmetric GUTs in the Conclusions.} 
in which the Higgs multiplets contain the electroweak triplet(s) $\phi$
because after the GUT gauge symmetry is broken the coupling 
$H^\dagger \phi H$ with the coefficient of order of the GUT scale is 
generated. We therefore reconsider the simplest $SU(5)$ GUT and show
by explicit calculation that this is indeed the case. Taking for the
input observables the Fermi constant $G_F$ and $\alpha_{\rm EM}$ 
we compute one loop corrections to the simplest electroweak observables
$\rho_{\rm low}$, $M_W$ and $M_Z$. Since the tree level expressions for 
$\rho_{\rm low}$ and $M_Z$ explicitly depend on the VEV of the electroweak 
triplet, tadpole contributions to these quantities do not cancel out and, 
as in the triplet model, do give extra contribution which is not suppressed 
by the inverse powers of the GUT scale. Unless the remaining free parameters 
of the model are artificially fine-tuned so to cancel (order by order) 
this contribution, being $\sim\ln(M_{\rm GUT}/M_Z)$ is 
unacceptably large. 

\section{The model and its spectrum}

We consider the simplest nonsupersymmetric $SU(5)$ GUT with fermions
in $\mathbf{10}$ and $\overline\mathbf{5}$ representations. The potential 
for the Higgs fields $\Phi$ transforming as $\mathbf{24}$ and $H$ 
transforming as $\mathbf{5}$ is
\begin{eqnarray}
{\cal V}\!\!&=&\!\!
-m^2_\phi{\rm tr}(\Phi^2)+{1\over4}a\left[{\rm tr}(\Phi^2)\right]^2
+{1\over2}b~{\rm tr}(\Phi^4)\nonumber\\
&&-m^2_HH^\dagger H+{\lambda\over4}\left(H^\dagger H\right)^2
+\alpha H^\dagger H~{\rm tr}(\Phi^2)+\beta H^\dagger\Phi^2 H
\label{eqn:potential}
\end{eqnarray}
Under $SU(3)\times SU(2)_L\times U(1)_Y$ the multiplet $H$ decomposes into an
$Y={1\over2}$ doublet of $SU(2)_L$ and an $SU(3)$ triplet ${\cal H}$ and the 
multiplet $\Phi$ contains a colour octet $\Phi_G$, two colour triplets 
$\Phi_X$ and $\Phi_Y$ forming an $SU(2)_L$ doublet, an $SU(2)_L$ triplet 
$\phi$ and a singlet $\varphi$:
\begin{eqnarray}
\Phi=\phi^a_G T^a_{\rm su3}
+ \phi_X^a T^a_X+ \phi_Y^a T^a_Y+ \phi^a T^a_{\rm su2}+ \varphi T^Y\nonumber
\equiv\Phi_G+\Phi_X+\Phi_Y+\phi+ \varphi T^Y\nonumber
\end{eqnarray}
It is clear that for $\langle\Phi\rangle\propto VT^Y$ the term proportional 
to $\beta$ in (\ref{eqn:potential}) generates the dangerous coupling 
$H^\dagger_{\rm SM}\phi H_{\rm SM}$ with the dimensionfull coefficient 
$\propto V$.

The equations determining the VEVs
\begin{eqnarray}
\langle\Phi\rangle=-3\sqrt{5\over3}VT^Y+v_\phi T^3_{\rm su2}\phantom{aaaaa}
\langle H_i\rangle={v_H\over\sqrt2}\delta_{i5}\nonumber
\end{eqnarray}
where $v_\phi$ is interpreted as the VEV of the $SU(2)_L$ triplet $\phi$ and
 $v_H$ is the VEV of the $SU(2)_L$ Higgs doublet read
\begin{eqnarray}
0\!\!&=&\!\!
-m^2_\phi15V+{a\over4}(225V^3+15Vv_\phi^2)+{b\over4}(105V^3+27Vv_\phi^2)
\nonumber\\
&\phantom{a}&
+{\alpha\over2}15Vv_H^2+{3\over4}\beta v_H^2(3V+v_\phi)\nonumber\\
0\!\!&=&\!\!
-m^2_H+{\lambda\over4}v^2_H+{\alpha\over2}(15V^2+v^2_\phi)
+{\beta\over4}(9V^2+6Vv_\phi+v^2_\phi)\label{eqn:eqforminimum}\\
0\!\!&=&\!\!-m^2_\phi v_\phi+{a\over4}(15V^2+v^2_\phi)v_\phi
+{b\over4}(27V^2+v^2_\phi)v_\phi+{\alpha\over2}v_H^2v_\phi+
{\beta\over4}v^2_H(3V+v_\phi)\nonumber
\end{eqnarray}
The hierarchy problem is easily seen: an extremely precise fine-tuning in 
the middle equation (\ref{eqn:eqforminimum}) is necessary in order to have
$v_H\ll V$. If $-m_H^2$ is of order $v_H^2\sim G_F^{-{1\over2}}$ then this 
requires $10\alpha+3\beta\sim v^2_H/V^2$. However it may well be that 
$-m_H^2$ and $V^2$ are of the same order and cancel against each other 
leading to $v_H^2\sim G_F^{-{1\over2}}$. Several useful relations necessary 
to analyse the spectrum of scalars can be derived from the equations 
(\ref{eqn:eqforminimum}). In particular, one can show 
that $v_\phi\sim v^2_H/V\ll v_H$.

The masses of the superheavy gauge bosons $X$ and $Y$ defined by the 
decomposition $A^a_\mu T^a_{\rm su5}= G^a_\mu T^a_{\rm su3}
+ X_\mu^a T^a_X+ Y_\mu^a T^a_Y+ W^a_\mu T^a_{\rm su2}+ B_\mu T^Y$
are then given by 
\begin{eqnarray}
M^2_X ={1\over4}g^2(5V-v_\phi)^2,\phantom{aaaa}
M^2_Y ={1\over4}g^2(5V+v_\phi)^2+{1\over4}g^2v_H^2\nonumber
\end{eqnarray}
and the masses of the electroweak gauge bosons are
\begin{eqnarray}
M^2_W={1\over4}g^2(v^2_H+4v_\phi^2)~,\phantom{aaaa}
M^2_Z={1\over4}{g^2\over c_W^2}v^2_H\nonumber
\end{eqnarray}
The $W$, $Z^0$ and $A^\gamma$ fields are defined as in the SM: 
$W^\pm={1\over\sqrt2}(W^1\mp iW^2)$, $Z^0=c_WW^3-s_WB$ and 
$A^\gamma=s_WW^3+c_WB$ except that $s_W^2=3/8$. 

The spectrum of scalars is as follows. The colour octet $\phi_G$ mass is
\begin{eqnarray}
M^2_{\phi_G}=-m^2_\phi+{1\over4}a(15V^2+v^2_\phi)+3bV^2+{1\over2}\alpha v^2_H
\end{eqnarray}
The Higgs triplet ${\cal H}$ 
originating from $H$ mixes with the triplet $\Phi_Y$
\begin{eqnarray}
\Phi_Y=c_Y ~G_Y+s_Y ~{\cal H}_Y\phantom{a}
\phantom{aaaaa}
{\cal H}=-s_Y G_Y~+c_Y ~{\cal H}_Y\label{eqn:HYGYdef}
\end{eqnarray}
where 
\begin{eqnarray}
s_Y={v_H\over\sqrt{(5V+v_\phi)^2+v_H^2}}~,
\phantom{aaaaa}
c_Y={5V+v_\phi\over\sqrt{(5V+v_\phi)^2+v_H^2}}\nonumber
\end{eqnarray}
$G_Y$ and $\Phi_X$ are massless and become the longitudinal components of the
vector bosons $Y$ and $X$, respectively. ${\cal H}_Y$ is massive with the mass
\begin{eqnarray}
M^2_{{\cal H}_Y}=-{\beta\over4}{V+v_\phi\over5V+v_\phi}[(5V+v_\phi)^2+v_H^2]
\end{eqnarray}
(it follows that $\beta<0$). The charged components $\phi^\pm$ of the 
electroweak triplet mix with the charged component $G^\pm$ of the doublet 
\begin{eqnarray}
G^+=c_\delta ~G^+_W+s_\delta ~H^+\phantom{aaaaaaaa}
\phi^+=-s_\delta ~G^+_W+c_\delta ~H^+\label{eqn:HGWdefs}
\end{eqnarray}
where 
\begin{eqnarray}
s_\delta=2{v_\phi\over v_H}{1\over\sqrt{1+4v^2_\phi/v^2_H}}~,
\phantom{aaaaa}
c_\delta={1\over\sqrt{1+4v^2_\phi/v^2_H}}~,\nonumber
\end{eqnarray}
$G_W$ and the $G^0$ component of the electroweak doublet are massless and
become the longitudinal components of the electroweak gauge bosons.
$H^\pm$ are massive with the mass 
\begin{eqnarray}
M^2_{H^\pm}=-{3\over4}\beta{V\over v_\phi}(v_H^2+4v^2_\phi)
\end{eqnarray}

Finally the matrix ${\cal M}^2$ of the masses squared of the three neutral 
scalars: the singlet $\varphi$, the neutral component 
$\phi_3$ of the triplet and $h$ coming from the doublet  has the structure
\begin{eqnarray}
{\cal M}^2=\left(\matrix{
\sim V^2&\sim v^2_H&\sim Vv_H\cr
\sim v^2_H&\sim V^2&\sim Vv_H\cr
\sim Vv_H&\sim Vv_H&{1\over2}\lambda v_H^2}\right)
\end{eqnarray}
As a result, the hierarchy of the off diagonal entries of the diagonalizing 
orthogonal matrix $O_{ij}$ is as follows:\footnote{For 
$10\alpha+3\beta\sim v^2_H/V^2$ the elements ${\cal M}^2_{13}$
and ${\cal M}^2_{31}$ would be of order $v^3_H/V$ and 
$O_{13}$, $O_{31}$ would vanish as $v^3_H/V^3$.}
\begin{eqnarray}
O_{12},O_{21}\sim{v_H^2\over V^2},\phantom{aaa} 
O_{13},O_{31},O_{23},O_{32}\sim{v_H\over V}
\end{eqnarray} 
We call the mass eigenstates $h_i^0\equiv(A^0,B^0,h^0)$. 
The SM physical Higgs boson is identified with $h^0$. In the limit 
$v_H/V\rightarrow0$ the masses $M^2_{A^0}\rightarrow{\cal M}^2_{11}$,
$M^2_{B^0}\rightarrow {\cal M}^2_{22}$, but the mass 
$M^2_{h^0}$ differs from ${\cal M}^2_{33}$ by a finite piece due to 
$\sim Vv_H$ off diagonal entries of ${\cal M}^2$. In calculations whose 
results we present in this paper it is important that 
\begin{eqnarray}
{\cal M}^2_{22}={1\over2}(a+b)v^2_\phi-{3\over4}\beta{Vv^2_H\over v_\phi}
\end{eqnarray} 
so that in this limit $M^2_{B^0}\rightarrow M^2_{H^\pm}$.

\section{Renormalization scheme}
\label {sec:scheme}

To compare at one loop predictions of the $SU(5)$ GUT with those of the SM 
we adopt the approach discussed in \cite{CHPOWA1,CHPOWA2}. Working in the
minimal subtraction scheme we trade the running GUT gauge coupling constant 
$g$ and the doublet VEV $v_H$ for the physical fine structure constant 
$\alpha_{\rm EM}$ and the Fermi constant $G_F$: 
\begin{eqnarray}
g^2&=&{4\pi\alpha_{\rm EM}\over s^2_W}
\left(1-{\delta\alpha\over\alpha}\right)
\nonumber\\
v^2_H &=& {1\over\sqrt2G_F}\left(1-4\sqrt2G_Fv^2_\phi+\Delta_G\right)
\label{eqn:basicrelsforschemeB}
\end{eqnarray}
where $\delta\alpha$ and $\Delta_G$ are the one-loop corrections:
\begin{eqnarray}
\alpha_{\rm EM}={g^2 s^2_W\over4\pi}+\delta\alpha,\phantom{aaaaa}
\sqrt2G_F={1\over v^2_H+4v^2_\phi}(1+\Delta_G)\nonumber
\end{eqnarray}
Since in the electroweak sector of the  GUT model there is one parameter less
than in the SM ($s_W$ is not a free parameter) both $M_W$ and $M_Z$ are 
calculable in terms of $G_F$, $\alpha_{\rm EM}$ and the remaining (running)
parameters. At the tree level one has
\begin{eqnarray}
&&M_W^2={\pi\alpha_{\rm EM}\over\sqrt2G_Fs^2_W}\nonumber\\
&&M_Z^2={\pi\alpha_{\rm EM}\over\sqrt2G_Fs^2_Wc^2_W}(1-4\sqrt2G_Fv^2_\phi)
\label{eqn:treelevelobservables}\\
&&\rho_{\rm low}={1\over1-4\sqrt2G_Fv^2_\phi}\nonumber
\end{eqnarray}
where $\rho_{\rm low}$ is defined as in \cite{CHPOWA1} by the ratio of the 
neutral to charged currents in the effective low energy weak interaction 
Lagrangian.

The correction $\delta\alpha$ is found using the technique used in 
\cite{CHPOWA1,CHPOWA2} (one can also use the method of \cite{WEINB}):
\begin{eqnarray}
{\delta\alpha\over\alpha}=
{g^2s^2_W\over16\pi^2}\left\{\left({2\over3}-7\ln{M^2_W\over Q^2}\right)+
{4\over3}\sum_fN_c^fQ_f^2\ln{m^2_f\over Q^2}\phantom{aaaaaaaaaaaaaa}\right.
\nonumber\\
+3\left(-{4\over3}\right)^2\left({2\over3}-7\ln{M^2_X\over Q^2}\right)
+3\left(-{1\over3}\right)^2\left({2\over3}-7\ln{M^2_Y\over Q^2}\right)
\nonumber\\
\left.+{1\over3}\left[\ln{M^2_{H^+}\over Q^2}
+3\cdot\left(-{1\over3}\right)^2\ln{M^2_{{\cal H}_Y}\over Q^2}
\right]\right\}\phantom{aaaaaaaaaaaaaaa}\label{eqn:deltaalpha}\\
=\left({\delta\alpha\over\hat\alpha}\right)^{\rm SM}
-{g^2\over16\pi^2}{353\over24}\ln{M^2_X\over Q^2}
\phantom{aaaaaaaaaaaaaaaaaaaaaaaaaa}\nonumber\\
+{g^2\over16\pi^2}\left\{
{34\over24}+{1\over8}\ln{M^2_{H^\pm}\over M^2_X}
+{1\over24}\ln{M^2_{{\cal H}_Y}\over M^2_X}
\right\}+\dots\phantom{aaa}\nonumber
\end{eqnarray}
For $\Delta_G$ we find 
\begin{eqnarray}
\Delta_G=-{\Pi_{WW}(0)\over M^2_W}
+{g^2\over16\pi^2}\left\{6-4\ln{M_Z^2\over Q^2}
-6\ln{M_X^2\over Q^2}+\dots\right.\phantom{aaaaaaaaa}\nonumber\\
\left.+\left(1-5s_W^2 +{7-14s^2_W+10s^4_W\over2s^2_W}
\left[{s^2_W\over c^2_W}{M^2_W\over M^2_Z-M^2_W}\right]\right)
\ln{M^2_W\over M_Z^2}\right\}\label{eqn:DeltaG}
\end{eqnarray}
In (\ref{eqn:deltaalpha}) and (\ref{eqn:DeltaG}) the ellipses stand for 
terms suppressed in the limit $V\rightarrow\infty$. In this limit the term 
in the square bracket of (\ref{eqn:DeltaG}) converges to $1$ and $\Delta_G$ 
differs its SM form only by the term $-6\ln(M_X^2/Q^2)$ (which arises from 
the $X,Y$ gauge bosons contribution to the $e\nu W$ vertex and to the $e$ 
and $\nu$ self energies) and by the contributions of heavy particles to 
$\Pi_{WW}$ which is given by (\ref{eqn:GUTPiWW}) in the \ref{app:formulae}.

\section{The $W$ boson mass}

The first observable we consider is the $W$ boson mass. In the adopted 
scheme the one loop formula reads
\begin{eqnarray}
M^2_W={\pi\alpha_{\rm EM}\over\sqrt2G_Fs^2_W}
\left(1+{\hat\Pi_{WW}(M^2_W)\over M^2_W}+\Delta_G
-{\delta\alpha\over\alpha}\right)\label{eqn:GUTMw}
\end{eqnarray}
Tadpoles do not contribute to (\ref{eqn:GUTMw}). This is because in the 
scheme based of $G_F$ and $\alpha_{\rm EM}$ the tree level $M_W^2$ 
(\ref{eqn:treelevelobservables}) expressed in terms of the input 
observables $G_F$ and $\alpha_{\rm EM}$ is independent of the VEVs. Using 
(\ref{eqn:WWdiv}) and (\ref{eqn:deltaalpha}) it is easy to check that the 
above expression is independent of the renormalization scale (i.e. it is 
finite).

The GUT formula (\ref{eqn:GUTMw}) has to be compared with the SM prediction
for $M_W^2$ in the same scheme, i.e. with only $G_F$ and $\alpha_{\rm EM}$ 
used as the input observables:
\begin{eqnarray}
M^2_W={\pi\alpha_{\rm EM}\over\sqrt2G_F\hat s^2}
\left\{
1+\left({\hat\Pi_{WW}(M^2_W)\over M^2_W}-
{\hat\Pi_{WW}(0)\over M^2_W}\right)^{\rm SM}
-\left({\delta\alpha\over\alpha}\right)^{\rm SM}\right.
\phantom{aaa}\nonumber\\
\left.+{e^2\over16\pi^2\hat s^2}\left(6-4\ln{M_Z^2\over Q^2}
+{7-12\hat s^2_W\over2\hat s^2_W}\ln{M_W^2\over M_Z^2}
\right)\right\}\label{eqn:SMMw}
\end{eqnarray}
By itself the expression in the curly bracket in (\ref{eqn:SMMw}) is not 
independent of the renormalization scale $Q$. Nevertheless, the entire 
expression is (up to terms of higher order)  $Q$ independent if one takes 
into account that in this scheme $\hat s^2$ is a running parameter:
\begin{eqnarray}
{1\over\hat s^2(Q)}&=&{1\over\hat s^2(\tilde Q)}\left\{
1+{\hat g^2\over16\pi^2}\left({109\over24}\right)
\ln{\tilde Q^2\over Q^2}\right\}\label{eqn:runningsinus}
\end{eqnarray}

Decoupling of heavy particle contributions is not manifest in the GUT formula 
(\ref{eqn:GUTMw}) for $M_W^2$. To see it it is necessary to absorb 
nonvanishing contributions of heavy particles into a redefinition of 
$s_W^2$. To this end we set in (\ref{eqn:GUTMw}) $Q\sim M_Z$ and write
\begin{eqnarray}
{\hat\Pi_{WW}(M_W^2)-\hat\Pi_{WW}(0)\over M^2_W}
=\left({\hat\Pi_{WW}(M_W^2)-\hat\Pi_{WW}(0)\over M^2_W}\right)^{\rm SM}
+{g^2\over M^2_W}\Delta_{WW}
\end{eqnarray}
Analysing the formula (\ref{eqn:GUTPiWW}) in which $h_3^0$ is identified 
with $h^0_{\rm SM}$ one finds that for $V\rightarrow\infty$ 
\begin{eqnarray}
\Delta_{WW}
=&-&4\left[\tilde A(M_W^2,\hat M_{H^\pm},\hat M_B)
-\tilde A(0,\hat M_{H^\pm},\hat M_B)\right]
\nonumber\\
&-&18\left[\tilde A(M_W^2,\hat M_Y,\hat M_X)
-\tilde A(0,\hat M_Y,\hat M_X)\right]
\nonumber\\
&-&6M^2_Wb_0(M_W^2,\hat M_X,\hat M_Y)+{M_W^2\over16\pi^2}
\nonumber
\end{eqnarray}
Since in the limit $\hat M_{H^\pm}^2-\hat M_B^2\sim v^2_H$ and
$\hat M_Y^2-\hat M_X^2\sim v^2_H$ we find
\begin{eqnarray}
16\pi^2\Delta_{WW}= \left({1\over3}+{3\over2}-6\right)\ln{M_X^2\over Q^2}
+{1\over3}\ln{M^2_{H^\pm}\over M_X^2}+M_W^2+\dots
\label{eqn:DeltaWW}
\end{eqnarray}
where the dots stand for  terms vanishing in the limit.

Combining together (\ref{eqn:deltaalpha}), (\ref{eqn:DeltaG}) and 
(\ref{eqn:DeltaWW}) to one loop accuracy  we get
\begin{eqnarray}
M^2_W={\pi\alpha_{\rm EM}\over\sqrt2G_Fs^2_W}
\left\{1+{g^2\over16\pi^2}\left({109\over24}\right)
\ln{M_X^2\over Q^2}+\Delta_T\right\}
\left\{\matrix{\phantom{a}\cr\phantom{a}}\right\}_{\rm SM}+\dots
\end{eqnarray}
where the content of the second curly bracket marked ``SM'' is the same as in 
(\ref{eqn:SMMw}). Comparison with (\ref{eqn:runningsinus})
shows that the first curly bracket changes the GUT value of $s_W^2$ into 
the value $\hat s^2(Q)$ of the SM at the renormalization scale $Q$. The 
finite threshold correction $\Delta_T$ reads
\begin{eqnarray}
\Delta_T=-{g^2\over16\pi^2}\left[{5\over12}
-{5\over24}\ln{M^2_{H^\pm}\over M^2_X}
+{1\over24}\ln{M^2_{{\cal H}_Y}\over M^2_X}\right]\label{eqn:DeltaT}
\end{eqnarray}
Thus, in the scheme with $G_F$ and $\alpha_{\rm EM}$ as the input observables,
$M_W^2$ computed in the GUT model coincides with $M_W^2$ obtained in the 
SM provided $\hat s^2(Q)$ is as predicted by the coupling constant
unification using the renormalization 
group analysis (supplemented with the threshold correction $\Delta_T$).
Other effects of the superheavy particles in this particular 
observable are suppressed in the one-loop approximation.

\section{The parameter $\rho_{\rm low}$}

This will not be so in the parameter $\rho_{\rm low}$ which we analyse now.
The precise definition of $\rho_{\rm low}$ we use is as in \cite{CHPOWA1}.
Performing the standard calculations we get the one-loop formula
\begin{eqnarray}
\rho_{\rm low} = {1\over1-4\sqrt2G_Fv^2_\phi}\left\{
1-{\Pi_{ZZ}(0)\over M^2_Z}
-{4g^2c_W^2\over16\pi^2}\left(\ln{M^2_W\over Q^2}
+3\ln{M^2_Y\over Q^2}\right)
-{\Delta_G\over1-4\sqrt2G_Fv^2_\phi}\right\}\nonumber
\end{eqnarray}
(we have omitted the contribution of the box diagrams to the process
$e^-\nu_\mu\rightarrow e^-\nu_\mu$ as the contribution of the additional
gauge bosons obviously vanishes in the limit $V\rightarrow\infty$).
Since in the lowest order
$(1-4\sqrt2G_F\hat v^2_\phi)\hat M^2_W=\hat M^2_Zc^2_W$, this can be 
rewritten as
\begin{eqnarray}
\rho_{\rm low} = {1\over1-4\sqrt2G_F\hat v^2_\phi}\left\{1
+{\Pi_{WW}(0)\over M^2_Zc^2_W}-{\Pi_{ZZ}(0)\over M^2_Z}
-{12g^2c_W^2\over16\pi^2}\ln{M^2_Y\over Q^2}\phantom{aaa}\right.\nonumber\\
\left.-{1\over1-4\sqrt2G_F v^2_\phi}
\left(-{6\over16\pi^2}g^2\ln{M^2_X\over Q^2}\right)+\dots\right\}
\label{eqn:formulaforrho}
\end{eqnarray}
where we have omitted terms which in the limit $V\rightarrow\infty$ reproduce 
the corresponding SM contributions. The full one-loop GUT expression for
$\Pi_{ZZ}$ is given in \ref{app:formulae}.

Careful analysis of the 1-PI contributions of heavy particles to 
(\ref{eqn:formulaforrho}) shows that all the dangerous terms cancel out 
leaving in the limit $V\rightarrow\infty$ only the contributions which arise 
from the SM. Similar cancellation of all dangerous 1-PI contributions was
also observed in \cite{SENSOK}. Yet, there is still the contribution of 
tadpoles to $\Pi_{WW}$ and $\Pi_{ZZ}$ (missed in \cite{SENSOK}) which we 
now analyse. In the SM tadpoles exactly cancel out in the combination
\begin{eqnarray}
{\Pi_{WW}\over M^2_W}-{\Pi_{ZZ}\over M^2_Z}\nonumber
\end{eqnarray}
which enters SM predictions for electroweak observables.
The more profound reason for their cancellation is, similarly as in the
case of (\ref{eqn:GUTMw}), the absence of $v_H$ in the tree-level formula 
for $\rho_{\rm low}$ (which in the SM is just 1), $M_W$ etc. after these
observables are 
expressed in terms of the input observables $G_F$ and $\alpha_{\rm EM}$. 
In the case considered here, $\rho_{\rm low}$ does depend at the tree level
on $v_\phi$ and, consequently, the tadpoles (\ref{eqn:tadpolesgeneral}) 
do not cancel out. Their contribution is (see \ref{app:tadpoles})
\begin{eqnarray}
\rho_{\rm low}={1\over1-4\sqrt2G_F\hat v^2_\phi}\left\{
1-{\hat g^2\over c^2_WM^2_Z}2v_\phi O_{2i}{{\cal T}_{h_i}\over M^2_{h_i}}
+\dots\right\}\label{eqn:tadpolestorho}
\end{eqnarray}
and represents direct corrections to the VEV of the neutral component 
$\phi^0$ of the electroweak triplet 
appearing in the prefactor of (\ref{eqn:formulaforrho}). Indeed, 
\begin{eqnarray}
{1\over1-4\sqrt2G_F(\hat v^2_\phi+\Delta v_\phi)^2}
={1\over1-4\sqrt2G_F\hat v^2_\phi}
\left\{1+{4\sqrt2G_F\over1-4\sqrt2G_F\hat v^2_\phi}
2\hat v_\phi\Delta v_\phi+\dots\right\}\nonumber
\end{eqnarray}
with 
\begin{eqnarray}
\Delta v_\phi=\langle\phi^0\rangle = O_{2i}\langle h_i\rangle=
O_{2i}{i\over-M^2_{h_i}}\left(-i{\cal T}_{h_i}\right)\label{eqn:Deltavphi}
\end{eqnarray}
exactly reproduces the contribution (\ref{eqn:tadpolestorho}). The 
contribution of tadpoles is also necessary to make (\ref{eqn:formulaforrho})
independent (up to two loop terms) of the renormalization scale $Q$
\cite{CHWA}. We do not attempt to check this explicitly because the 
derivation of the necessary renormalization group equation for $v_\phi^2$
appears cumbersome, but in \cite{CHPOWA1} and \cite{CHPOWA2} we have managed 
to demonstrate this explicitly for the models considered in these papers. 

The tadpoles ${\cal T}_{h_i}$ have dimension (mass)$^3$. Hence, as 
$V\rightarrow\infty$, they can grow at most as $V^3$ (up to logarithms). 
Since $v_\phi\sim1/V$ and $O_{21}\sim1/V^2$ the contribution of
${\cal T}_{h_1}$ is well suppressed. $O_{22}$ is of order $1$ but the 
leading, order $V^3$ terms in ${\cal T}_{h_2}$ happen to cancel out
and this contribution is also suppressed. However, 
it is a matter of a straightforward analysis (though the derivation of the 
necessary couplings is lengthy) to check that in ${\cal T}_{h_3}$ there are 
terms of order $V^2$ which, in view of the fact that $O_{23}$ vanishes only 
as $1/V$ (and $M_{h_3}$ is identified with the SM Higgs mass, of order 
$v_H$) give contributions to $\rho_{\rm low}$ growing logarithmically
for $V\rightarrow\infty$ (the formula for the dominant terms in 
${\cal T}_{h_3}$ is given in \ref{app:tadpoles}). Thus, $\rho_{\rm low}$
computed in the $SU(5)$ GUT differs from  $\rho_{\rm low}$ computed in 
SM by the term which can be interpreted as the custodial symmetry breaking,
logarithmically dependent on the unification scale $M_X$, correction to
$v_\phi$ in the tree level term.  The reason for this is that once 
the $SU(2)_L\times U(1)_Y$ symmetry is broken, the triplet is not
protected against acquiring radiatively generated VEV and since there is
a dimensionfull, $\sim V$, coupling $H^\dagger\phi H$, the effects of 
this correction to $\langle\phi\rangle$ is not suppressed in electroweak
observables.

\section{The $Z^0$ boson mass}

The same nondecoupling of tadpole contribution occurs in the GUT formula 
for $M_Z$. At one loop one finds
\begin{eqnarray}
M_Z^2={\pi\alpha_{\rm EM}\over\sqrt2G_Fs^2_Wc^2_W}
(1-4\sqrt2G_F\hat v^2_\phi)
\left\{1-{\delta\alpha\over\alpha}
+{\Delta_G\over1-4\sqrt2G_F\hat v^2_\phi}+{\Pi_{ZZ}(M^2_Z)\over M^2_Z}
\right\}\label{eqn:GUTZbosnmass}
\end{eqnarray}
Checking the renormalization scale independence of this formula would be 
quite complicated. It is however straightforward to check it assuming that 
the formula (\ref{eqn:formulaforrho}) for $\rho_{\rm low}$ is $Q$ independent. 

After a careful analysis of the 1-PI contribution (\ref{eqn:GUTZbosnmass}) 
can be rewritten in the form 
\begin{eqnarray}
M_Z^2={\pi\alpha_{\rm EM}\over\sqrt2G_Fs^2_Wc^2_W}
(1-4\sqrt2G_F\hat v^2_\phi)
\left\{1+{g^2\over16\pi^2}\left(1-{s^2_W\over c^2_W}\right)
\left({109\over24}\ln{\hat M^2_X\over Q^2}+\Delta_T\right)\right\}
\nonumber\\
\times\left\{1+\left({\Pi_{ZZ}(M^2_Z)\over M^2_Z}
-{\Pi_{WW}(0)\over c_W^2M^2_Z}\right)^{\rm1-PI}_{\rm SM}
-\left({\delta\alpha\over\alpha}\right)_{\rm SM}\right.
\phantom{aaaaaaaaaaaaa}\label{eqn:limGUTZbosnmass}\\
\left.-{4g^2\over16\pi^2}\ln{\hat M^2_Z\over Q^2}
+{\hat g^2\over c^2_WM^2_Z}2v_\phi O_{2i}{{\cal T}_{h_i}\over M^2_{h_i}}
+\dots\right\}\phantom{aaaaaaa}\nonumber
\end{eqnarray}
where $\Delta_T$ is exactly the same as in (\ref{eqn:DeltaT}) and the ellipses
stand for terms vanishing for $V\rightarrow\infty$. The first curly bracket 
converts $s^2_Wc^2_W$ in front of (\ref{eqn:limGUTZbosnmass}) into the 
running parameters $\hat s^2(Q)\hat c^2(Q)$ and the rest of the result would 
be just as in the SM (in the same scheme with $G_F$ and $\alpha_{\rm EM}$ 
taken as the input observables) if there were no tadpoles. Again
the tadpoles can be interpreted as the loop correction to the tree
level VEV $v_\phi$ of the triplet in the factor in front of the formula
for $M_Z$. 

As we have argued in \cite{CHPOWA2}, the contribution of tadpoles to 
electroweak observables can be summarized by their contribution to the 
Peskin-Takeuchi $T$ parameter \cite{PETA}:
\begin{eqnarray}
\alpha_{\rm EM}\Delta T=4\sqrt2G_F2v_\phi\Delta v_\phi
\end{eqnarray}
where $\Delta v_\phi$ is given by (\ref{eqn:Deltavphi}). Since 
$\Delta v_\phi\sim\ln M_X$ where $M_X\sim10^{14}$ GeV this is an unacceptably 
large contribution unless the parameters are artificially tuned so as to
cancel $\Delta v_\phi$.

\section{Conclusions}

We have confirmed by direct calculation that the violation of the 
Appelquist-Carazzone decoupling of heavy states found in \cite{CHPOWA2} 
holds also in the simplest nonsupersymmetric GUT model. To this end we 
have applied the renormalization scheme based on two input observables
$\alpha_{\rm EM}$, $G_F$ and compared the formulae for $M_W$, $\rho_{\rm low}$
and $M_Z$ obtained in the GUT model with the SM formulae obtained using the 
same renormalization scheme (as argued in \cite{CHPOWA1,CHPOWA2} this is 
the right procedure for checking whether the Appelquist-Carazzone decoupling 
holds). As in \cite{CHPOWA2} the nondecoupling is due to the radiative 
generation of the electroweak triplet VEV which is not suppressed because 
of the growing with $V$ effective coupling of the SM doublet to the triplet. 
Although in principle one can chose the parameters of the model so to cancel 
the tadpoles (at a given order of perturbation expansion) this requires an 
additional severe fine-tuning. This additional fine-tuning comes on the 
top of the well known hierarchy problem but it is conceptually slightly 
different from the latter. As we have explained in the introduction, and 
showed by explicit calculation, the fine tuning necessary to separate the 
electroweak scale from the GUT scale is not seen once $G_F$ is taken for 
one of the input observables. In contrast, the new fine-tuning is 
``more physical'' because it is seen even after expressing the electroweak 
observables in terms of $G_F$ and $\alpha_{\rm EM}$.

The additional fine-tuning identified in this paper appears important in 
view of the recent revival of nonsupersymetric GUTs \cite{DORS,MOSEN} 
and nonsupersymmetric models of neutrino masses \cite{BASEN} employing 
electroweak Higgs triplets (so called see-saw mechanism of type II 
\cite{MOSEN}).

From the theoretical perspective, nondecoupling found in \cite{CHPOWA2} and 
in this paper constitutes an interesting counterexample to the conjecture 
put forward by Senjanovic, Sokorac and Wilczek \cite{SENSOK} that decoupling 
in the full theory (with both VEVs taken into account) is ensured by the 
gauge invariance of the effective theory obtained by eliminating heavy 
degrees of freedom for $v_H=0$. It turns out that in the presence of the 
effective dimensionfull couplings proportional to the larger mass scale this 
conjecture is not true. This has consequences for the common practice of 
computing the $S$, $T$, $U$ parameters by adding to the SM Lagrangian higher 
dimension $SU(2)_L\times U(1)_Y$ invariant operators whose coefficients are 
determined before taking into account the electroweak symmetry breaking. For 
example, in \cite{GRSKTE} the authors conclude that the electroweak triplet 
contribution to the $T$ parameter is of order $\sim\mu^2v^2_H/m^4_\phi$ 
(where $\mu$ is the dimensionfull coupling of the triplet to doublets
and $m_\phi^2$ is the Lagrangian triplet mass parameter) which 
corresponds precisely to the contribution of the tree level triplet VEV 
$v_\phi$. However, as follows from the analysis performed in \cite{CHPOWA2} 
for the natural choice $\mu\sim m_\phi$ (and other parameters of the model 
generic) the results of the full calculations can be reproduced only if the 
$\sim\ln m_\phi$ contribution of tadpoles is included in the $T$ parameter.

Finally, we anticipate that this effect should not be present in 
supersymmetric GUT in which there is a mutual supersymmetric cancellation 
of bosonic and fermionic tadpoles. Violation of the Appelquist-Carazzone 
decoupling found in \cite{CHPOWA2} and in this paper can be therefore
considered yet another argument for supersymmetry in the context of GUTs.

\section*{Acknowledgments}

P.H.Ch. was partially supported by the Polish KBN grants 1 P03B 108 30 
(for years 2006--2008) and 1 P03B 099 29 (for years 2005--2007) and by 
the European Community Contract MTKD-CT-2005-029466 - project "Particle 
Physics and Cosmology: the Interface" (for years 2006-2010).

\newpage
\renewcommand{\thesection}{Appendix~\Alph{section}}
\renewcommand{\theequation}{\Alph{section}.\arabic{equation}}
\setcounter{section}{0}

\section{Vector bosons self energies}
\label{app:formulae}
\setcounter{equation}{0}

One particle irreducible (1-PI) contribution to the $W$-boson self energy
$\Pi_{WW}(q^2)$ is
\begin{eqnarray}
&&{1\over2}g^2\sum_{k=1}^3\left[4\tilde A(q^2,m_{e_k},0)+(q^2-m^2_{e_k})
b_0(q^2,m_{e_k},0)\right]\nonumber\\
&+&{3\over2}g^2\sum_{k,l=1}^3|V_{CKM}^{kl}|^2
\left[4\tilde A(q^2,m_{u_k},m_{d_l})
+(q^2-m^2_{u_k}-m^2_{d_l})b_0(q^2,m_{u_k},m_{d_l})\right]
\nonumber\\
&-&g^2(O_{3i}c_\delta+2O_{2i}s_\delta)^2\tilde A(q^2,M_W,M_{h_i})
\nonumber\\
&-&g^2(O_{3i}s_\delta-2O_{2i}c_\delta)^2\tilde A(q^2,M_{H^\pm},M_{h_i})
\nonumber\\
&-&6g^2g^2c^2_Y\tilde A(q^2,M_Y,M_X)
\nonumber\\
&-&6g^2s^2_Y\tilde A(q^2,M_{{\cal H}_Y},M_X)
\nonumber\\
&-&g^2c^2_\delta\tilde A(q^2,M_W,M_Z)
\nonumber\\
&-&g^2s^2_\delta\tilde A(q^2,M_{H^\pm},M_Z)
\nonumber\\
&+&{1\over4}g^4(v_HO_{3i}+4v_\phi O_{2i})^2
b_0(q^2,M_W,M_{h_i})\nonumber\\
&+&g^2s^2_WM^2_Wb_0(q^2,M_W,0)\nonumber\\
&+&{g^4s^2_W\over4(v_H^2+4v_\phi^2)}
\left({s_W\over c_W}v_H^2-4{c_W\over s_W}v_\phi^2\right)^2
b_0(q^2,M_W,M_Z)\label{eqn:GUTPiWW}\\
&+&{g^4v_H^2v_\phi^2\over c^2_W(v_H^2+4v_\phi^2)}
b_0(q^2,M_{H^\pm},M_Z)\nonumber\\
&+&{3\over2}g^4\left({5V+3v_\phi\over2}\right)^2b_0(q^2,M_Y,M_X)
\nonumber\\
&+&{3\over2}g^4{[(5V-3v_\phi)(5V+v_\phi)-v_H^2]^2\over
4[(5V+v_\phi)^2+v_H^2]}b_0(q^2,M_Y,M_X)
\nonumber\\
&+&{3\over2}g^4{(5V-v_\phi)^2v_H^2\over[(5V+v_\phi)^2+v_H^2]}
b_0(q^2,M_{{\cal H}_Y},M_X)
\nonumber\\
&-&g^2s^2_W\left[8\tilde A(q^2,M_W,0)
+(4q^2+M^2_W)b_0(q^2,M_W,0)-{1\over16\pi^2}{2\over3}q^2\right]\nonumber\\
&-&g^2c^2_W\left[8\tilde A(q^2,M_W,M_Z)
+(4q^2+M^2_W+M^2_Z)b_0(q^2,M_W,M_Z)
-{1\over16\pi^2}{2\over3}q^2\right]\nonumber\\
&-&{3\over2}g^2\left[8\tilde A(q^2,M_X,M_Y)
+(4q^2+M^2_X+M^2_Y)b_0(q^2,M_X,M_Y)-{1\over16\pi^2}{2\over3}q^2\right]\nonumber
\end{eqnarray}
The first two lines are the contributions of fermions and the 
successive come from loops of
$G^\pm_Wh_i$, $H^\pm h_i$, $\Phi_XG_Y$, $\Phi_X{\cal H}_Y$, $G^\pm_WG_Z$, 
$H^\pm G_Z$, $W^\pm h_i$, $G^\pm_W\gamma$, $G^\pm_WZ^0$, $H^\pm Z^0$, 
$\Phi_XY$,  $XG_Y$, $X{\cal H}_Y$, 
$W^\pm Z^0$, $W^\pm \gamma$ and $XY$. 

In analysing this expression it is useful to notice that the 
coefficients of the two terms with $b_0(q^2,M_Y,M_X)$ can be combined giving
\begin{eqnarray}
+{3\over2}g^2\left(M_Y^2+M^2_X-g^2v_H^2
+\dots\right)\label{eqn:combination}
\end{eqnarray}
where ellipses stand for terms suppressed for $V\rightarrow\infty$. 

The dependence of (\ref{eqn:GUTPiWW}) on the 
renormalization scale $Q$ is as follows
\begin{eqnarray}
[\Pi_{WW}(q^2)]={g^2\over16\pi^2}
\sum_{\rm gen}\left\{
\left({1\over3}q^2-{1\over2} m_e^2\right)
+ N_c\left({1\over3}q^2-{1\over2} m_u^2-{1\over2} m_d^2\right)
\right\}\ln{1\over Q^2}\nonumber\\
+{g^2\over16\pi^2}\left\{
\left(-{22\over3}q^2\right)
+{1\over4}g^2v_H^2\left({s^2_W\over c_W^2}-1\right)
+10g^2v_\phi^2\right\}\ln{1\over Q^2}\phantom{aaaaa}
\label{eqn:WWdiv}
\end{eqnarray}
\vskip0.3cm

The 1-PI contribution of fermions to the $Z^0$ boson self energy
$\Pi_{ZZ}(q^2)$ is as in the SM:
\begin{eqnarray}
&&{g^2\over2c_W^2}\sum_fN_c^fa_+^f\left[2\tilde A(q^2,m_f,m_f)
+({q^2\over2}-m^2_f)b_0(q^2,m_f,m_f)\right]\nonumber\\
&+&{g^2\over2c_W^2}\sum_fN_c^fa_-^fm^2_fb_0(q^2,m_f,m_f)
+3\left[\tilde A(q^2,0,0)+{q^2\over4}b_0(q^2,0,0)\right]\label{eqn:GUTPiZZfer}
\end{eqnarray}
where the sums in the two first terms extend to all fermions except neutrinos
(which are accounted for by the third term),
$a_+^f=1-4|Q_f|s^2_W+8|Q_f|^2s^4_W$, 
$a_-^f= -4|Q_f|s^2_W+8|Q_f|^2s^4_W$.
The bosonic 1-PI contribution to $\Pi_{ZZ}(q^2)$ (in the same units) reads
\begin{eqnarray}
&&-{g^2\over c_W^2}O_{3i}O_{3i}\tilde A(q^2,M_Z,M_{h_i}) \nonumber\\
&&-3_c\cdot g^2c^2_W\left(c_Y^2+{s^2_W\over3c^2_W}s_Y^2\right)^2
4\tilde A(q^2,M_Y,M_Y) \nonumber\\
&&-3_c\cdot g^2c^2_W\left(s_Y^2+{s^2_W\over3c^2_W}c_Y^2\right)^2
4\tilde A(q^2,M_{{\cal H}_Y},M_{{\cal H}_Y}) \nonumber\\
&&-3_c\cdot 2g^2c^2_Wc_Y^2s_Y^2\left(1-{s^2_W\over3c^2_W}\right)^2
4\tilde A(q^2,M_Y,M_{{\cal H}_Y}) \nonumber\\
&&-{g^2\over c_W^2}
((1-2s^2_W)c_\delta^2+2c_W^2s_\delta^2)^2\tilde A(q^2,M_W,M_W)
\nonumber\\
&&-{g^2\over c_W^2}((1-2s^2_W)s_\delta^2+2c_W^2c_\delta^2)^2
\tilde A(q^2,M_{H^+},M_{H^+})\nonumber\\
&&-2{g^2\over c_W^2}c_\delta^2s_\delta^2\tilde A(q^2,M_W,M_{H^+})\nonumber\\
&&+2{g^4s_W^2\over4(v_H^2+4v_\phi^2)}
\left({s_W\over c_W}v_H^2-{c_W\over s_W}4v_\phi^2\right)^2
b_0(q^2,M_W,M_W)\label{eqn:GUTPiZZbos}\\
&&+2{g^4v_H^2v_\phi^2\over c^2_W(v_H^2+4v_\phi^2)}
b_0(q^2,M_W,M_{H^+})\nonumber\\
&&+{g^2\over c_W^2}O_{3i}O_{3i}M^2_Zb_0(q^2,M_Z,M_{h_i})\nonumber\\
&&+2\cdot3_c\cdot{1\over4}g^4c_W^2
{[(5V+v_\phi)^2+v_H^2-(v_H^2/c^2_W)]^2\over
(5V+v_\phi)^2+v_H^2}b_0(q^2,M_Y,M_Y)\nonumber\\
&&+2\cdot3_c\cdot{g^4\over4c^2_W}{(5V+v_\phi)^2v_H^2\over
(5V+v_\phi)^2+v_H^2}b_0(q^2,M_Y,M_{{\cal H}_Y})\nonumber\\
&&-g^2c^2_W\left[8\tilde A(q^2,M_W,M_W)
+(4q^2+2M^2_W)b_0(q^2,M_W,M_W)
-{1\over16\pi^2}{2\over3}q^2\right]\nonumber\\
&&-3_c\cdot g^2c^2_W\left[8\tilde A(q^2,M_Y, M_Y)
+(4q^2+2M^2_Y)b_0(q^2,M_Y,M_Y)
-{1\over16\pi^2}{2\over3}q^2\right]
\nonumber
\end{eqnarray}
The successive lines are the contributions of:
$G^0h_i$, $G_YG_Y$, ${\cal H}_Y{\cal H}_Y$, ${\cal H}_YG_Y$,
$G^\pm_WG_W^\mp$, $H^\pm H^\mp$, $G_W^\pm H^\mp$,
$W^\pm G_W^\mp$, $W^\pm H^\mp$, $Z^0h_i$
$Z^0G_Y$, $Z^0{\cal H}_Y$, $W^\pm W^\mp$ and $YY$.

The dependence on the 
renormalization scale $Q$ of (\ref{eqn:GUTPiZZfer}) is: 
\begin{eqnarray}
\left[\hat\Pi_{ZZ}(q^2)\right]^{\rm (f)}
={1\over16\pi^2}{g^2\over 3c^2_W}q^2\sum_{\rm gen}\left[
1-2s^2_W+4s^4_W+N_c\left(1-2s^2_W+{20\over9}s^4_W\right)\right]
\ln{1\over Q^2}
\nonumber\\
-{1\over16\pi^2}{\hat g^2\over c^2_W}\sum_{\rm gen}\left({1\over2}m^2_e
+{N_c\over2}m^2_u+{N_c\over2}m^2_d\right)\ln{1\over Q^2}
\phantom{aaaaaaaaaaaaaa}
\end{eqnarray}
and that of (\ref{eqn:GUTPiZZbos}) reads
\begin{eqnarray}
\left[\hat\Pi_{ZZ}(q^2)\right]^{\rm (b)}
={1\over16\pi^2}{g^2\over c^2_W}q^2
\left\{{1\over2}+c_W^4+{1\over9}s_W^4-s_W^2
+{2\over3}s_W^4-{40\over3}c_W^4\right\}\ln{1\over Q^2}
\nonumber\\
+{1\over16\pi^2}{g^2\over c^2_W}\left\{2g^2(1-2c^2_W)v_H^2
+{1\over4}{g^2\over c_W^2}v_H^2\right\}\ln{1\over Q^2}\phantom{aaaa}
\end{eqnarray}

The standard functions $b_0(q^2,m_1,m_2)$ and $\tilde A(q^2,m_1,m_2)$ are 
defined for example in the Appendix of \cite{CHPOWA1}. For 
$q^2\ll$max$(m^2_1,m^2_2)$ we have 
\begin{eqnarray}
\tilde A(q^2,m_1,m_2)-\tilde A(0,m_1,m_2)={1\over16\pi^2}{q^2\over18}
-{q^2\over12}b_0(0,m_1,m_2)
+{q^2\over6}b_0^\prime(0,m_1,m_2)\nonumber\\
-{q^2\over24}(m_1^2-m_2^2)^2b_0^{\prime\prime}(0,m_1,m_2)+\dots
\phantom{aaaaa}\label{eqn:A1}
\end{eqnarray}
(primes mean derivatives with respect to $q^2$).
For $m_1^2-m_2^2\sim q^2$ this reduces to 
\begin{eqnarray}
\tilde A(q^2,m_1,m_2)-\tilde A(0,m_1,m_2)=
-{1\over16\pi^2}{q^2\over12}\ln{m^2_1\over Q^2}\dots\label{eqn:A2}
\end{eqnarray}

\section{Tadpoles}
\label{app:tadpoles}
\setcounter{equation}{0}

Tadpole contribution to $\Pi_{WW}$ and $\Pi_{ZZ}$ are
\begin{eqnarray}
&&i\hat\Pi_{WW}^{\rm Tad}={i\over2}\hat g^2(\hat v_HO_{3i}+4\hat v_\phi O_{2i})
{i\over-M^2_{h_i}}(-i{\cal T}_{h_i})\nonumber\\
&&i\hat\Pi_{ZZ}^{\rm Tad}={i\over2c^2_W}\hat g^2\hat v_HO_{3i}
{i\over-M^2_{h_i}}(-i{\cal T}_{h_i})\label{eqn:tadpolesgeneral}
\end{eqnarray}
where $-i{\cal T}_{h_i}$ is the sum of 1-PI diagrams with 
a single external line of the scalar $h_i$.

As analysed in the text, contribution of tadpoles ${\cal T}_{h_1}$ and
${\cal T}_{h_2}$ is suppressed for $V\rightarrow\infty$. We display only 
those contributions to ${\cal T}_{h_3}$, which cause the nondecoupling 
of the electroweak triplet that is, those which grow as $V^2$ 
(up to logarithms).

\noindent {\it Gauge bosons and ghosts}
\begin{eqnarray}
{\cal T}_{h_3}^{(X,Y)}={1\over16\pi^2}{45\over2}g^2{c_W\over s_W}VO_{13}\left\{
M_X^2\left(\ln{M^2_X\over Q^2}-{1\over3}\right)+
M_Y^2\left(\ln{M^2_Y\over Q^2}-{1\over3}\right)\right\}\nonumber\\
-{1\over16\pi^2}{9\over2}g^2v_HO_{33}M_Y^2\left(\ln{M^2_Y\over Q^2}-{1\over3}
\right)\phantom{aaaaaaaaaaaaaaaaaaaaaaa}\nonumber
\end{eqnarray}

\noindent {\it Goldstone bosons}\\
Contribution of $\Phi_X$:
\begin{eqnarray}
{\cal T}^{(\Phi_X)}_{h_3}=
{1\over16\pi^2}\left\{-\left({5\over2}a+{7\over6}b\right)
\sqrt{3\over5}VO_{13}-b\alpha ~v_H O_{33}+VO_{23}
\right\}M^2_X\left(\ln{M^2_X\over Q^2}-1\right)\nonumber
\end{eqnarray}
Contribution of $G_Y$:
\begin{eqnarray}
{\cal T}_{h_3}^{(G_Y)}=
{1\over16\pi^2}
\left\{-c_Y^2\left({5\over2}a+{7\over6}b\right)\sqrt{3\over5}VO_{13}
+bc_Y^2VO_{23}\right.\phantom{aaaaaaaaaaaaaaaa}\nonumber\\
\left.\phantom{aa}+\alpha ~v_Hc_Y^2O_{33}
+{\beta\over2}v_Hc_Y^2O_{33}
+{\beta\over2}Vc_Ys_Y O_{33}
\right\}M^2_Y\left(\ln{M^2_Y\over Q^2}-1\right)\nonumber
\end{eqnarray}

\noindent {\it Charged Higgs bosons}\\
Contribution of ${\cal H}_Y$:
\begin{eqnarray}
{\cal T}_{h_3}^{({\cal H}_Y)}=
{1\over16\pi^2}\left\{{\lambda\over2}v_H c_Y^2O_{33}
-c_Y^2\left(5\alpha +{2\over3}\beta\right)\sqrt{3\over5}VO_{13}
\right.\phantom{aaaaaaaaaaaaaaaa}\nonumber\\
\left.
-{\beta\over2}Vc_Ys_Y O_{33}\right\}
M^2_{{\cal H}_Y}\left(\ln{M^2_{{\cal H}_Y}\over Q^2}-1\right)\nonumber
\end{eqnarray}
Contribution of $H^\pm$:
\begin{eqnarray}
{\cal T}_{h_3}^{(H^\pm)}=
{1\over16\pi^2}\left\{-{1\over2} c_\delta^2
\left(5a+9b\right)\sqrt{3\over5}V~O_{13}
\right.\phantom{aaaaaaaaaaaaaaaaaaaaa}\nonumber\\
\left.+{1\over2}\beta v_Hc_\delta^2O_{33}
-3Vs_\delta c_\delta O_{33}\right\}
M^2_{H^\pm}\left(\ln{M^2_{H^\pm}\over Q^2}-1\right)\nonumber
\end{eqnarray}

\noindent {\it Neutral Higgs bosons}\\
\noindent Contribution of $\phi_G$:
\begin{eqnarray}
{\cal T}^{(\phi_G)}_{h_3}=
{1\over16\pi^2}{1\over2}\left\{-\left({5\over2}a+2b\right)\sqrt{3\over5}VO_{13}
+\alpha ~v_H O_{33} \right\}M^2_{\phi_G}
\left(\ln{M^2_{\phi_G}\over Q^2}-1\right)\nonumber
\end{eqnarray}
\noindent Contribution of $h_i^0$:
\begin{eqnarray}
{\cal T}^{(A,B)}_{h_3}\!\!\!\!\!\!\!\!
&&={1\over16\pi^2}\sum_{j=1}^2\left\{-{1\over4}\left(5a+{7\over3}b\right)
\sqrt{3\over5}V~O_{13}O_{1j}O_{1j}
-{1\over4}(5a+9b)V~O_{13}O_{2j}O_{2j}\right.\nonumber\\
&&\phantom{aaaaa}
+{1\over2}\alpha v_HO_{33}(O_{1j}O_{1j}+O_{2j}O_{2j})\nonumber\\
&&\phantom{aaaaa}\left.
+{1\over4}\beta v_H~O_{33}\left({3\over5}~O_{1j}O_{1j}
+O_{2i}O_{2j}\right)\right\}M^2_{h_j}\left(\ln{M^2_{h_j}\over Q^2}-1\right)
\nonumber\\
&&+{1\over16\pi^2}\sum_{j=1}^22\left\{-{3\over4}\sqrt{5\over3}
aV~O_{1j}O_{13}O_{1j}\right.\nonumber\\
&&\phantom{aaaaa}
+{1\over4}b\left(-{7\over3}\sqrt{3\over5}V~O_{13}O_{1j}O_{1j}
\right)-{3\over2}\sqrt{5\over3}\alpha V~O_{1j}O_{33}O_{3j}\nonumber\\
&&\phantom{aaaaa}\left.
-{3\over4}\beta V\left(\sqrt{3\over5}~O_{1j}-O_{2j}\right)
O_{33}O_{3j}\right\}M^2_{h_j}\left(\ln{M^2_{h_j}\over Q^2}-1\right)\nonumber
\end{eqnarray}
If the fine-tuning necessary to separate the GUT and the electroweak scales
requires $10\alpha+3\beta\sim v^2_H/V^2$, then the elements $O_{13}$ and 
$O_{31}$ are very small and all terms with $VO_{13}$ or $VO_{31}$ can be 
dropped. We have remarked however that this does not appear to be the only 
possibility. Still, even if $O_{13}$ and $O_{31}$ are more suppressed than 
we have assumed above, ${\cal T}_{h_3}$ grows as $V^2$ and the
decoupling is violated.

\end{document}